\documentclass[twocolumn,floatfix,aps,superscriptaddress]{revtex4}
\usepackage{graphicx}
\usepackage{amsmath}
\usepackage{amssymb}
\usepackage{bm}

\begin{document}

\title{Topologically protected Landau level in the vortex lattice of a Weyl superconductor}
\author{M. J. Pacholski}
\affiliation{Instituut-Lorentz, Universiteit Leiden, P.O. Box 9506, 2300 RA Leiden, The Netherlands}
\author{C. W. J. Beenakker}
\affiliation{Instituut-Lorentz, Universiteit Leiden, P.O. Box 9506, 2300 RA Leiden, The Netherlands}
\author{\.{I}. Adagideli}
\affiliation{Faculty of Engineering and Natural Sciences, Sabanci University, Orhanli-Tuzla, Istanbul, Turkey}
\date{October 2017}

\begin{abstract}
The question whether the mixed phase of a gapless superconductor can support a Landau level is a celebrated problem in the context of \textit{d}-wave superconductivity, with a negative answer: The scattering of the subgap excitations (massless Dirac fermions) by the vortex lattice obscures the Landau level quantization. Here we show that the same question has a positive answer for a Weyl superconductor: The chirality of the Weyl fermions protects the zeroth Landau level by means of a topological index theorem. As a result, the heat conductance parallel to the magnetic field has the universal value $G=\tfrac{1}{2}g_0 \Phi/\Phi_0$, with $\Phi$ the magnetic flux through the system, $\Phi_0$ the superconducting flux quantum, and $g_0$ the thermal conductance quantum. 
\end{abstract}
\maketitle

\emph{Introduction ---}
In 1998 Gor'kov, Schrieffer \cite{Gor98}, and Anderson \cite{And98} made the remarkable prediction that the excitation spectrum in the mixed phase of a high-$T_c$ superconductor (with massless quasiparticles at nodal points of the \textit{d}-wave pair potential) has the Landau levels of the relativistic Dirac equation. This was nearly a decade before the quantum Hall effect of massless electrons was measured in graphene \cite{Nov05,Zha05}, and it would have marked the first appearance in the solid state of a  magnetic-field independent zeroth Landau level. 

It did not turn out that way: The spatially varying supercurrent in the Abrikosov vortex lattice strongly scatters the quasiparticles \cite{Mel99}, even if the vortices overlap and produce a uniform magnetic field. Since Franz and Te\u{s}anovi\'{c} \cite{Fra00} we know that the quasiparticles in the mixed phase of a \textit{d}-wave superconductor retain the zero-field Dirac cone, the main effect of the magnetic field being a renormalization of the Fermi velocity \cite{Yas99,Mar00,Kop00,Mor00,Vaf01a,Kna01,Vaf01b,Vis01,Vaf06,Mel07}. Recent proposals \cite{Mas17,Nic17,Liu17} use strain to mimic the effect of a magnetic field in a \textit{d}-wave superconductor without breaking time-reversal symmetry, but the coexistence of Landau levels and a vortex lattice has remained elusive.

\begin{figure}[tb]
\centerline{\includegraphics[width=1\linewidth]{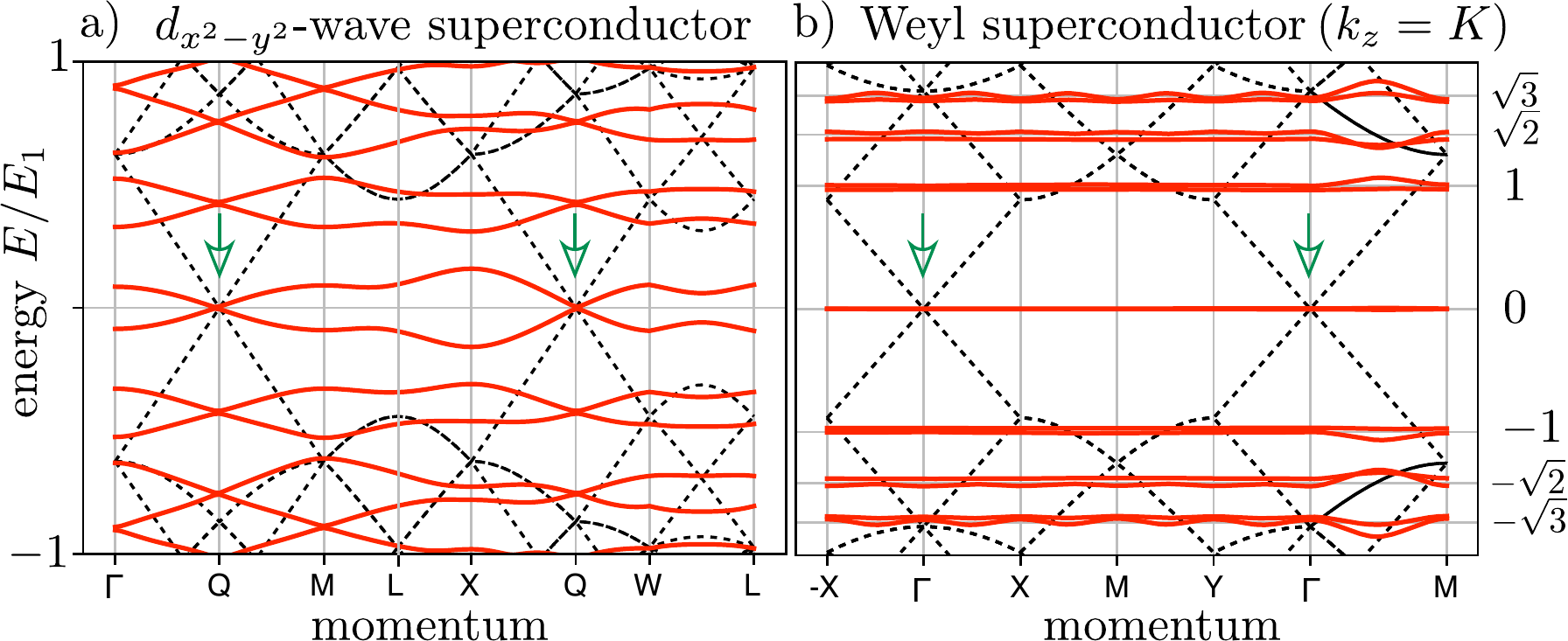}}
\caption{Excitation spectrum of a nodal superconductor in zero magnetic field (black dashed curves) and in the mixed phase with a square lattice of Abrikosov vortices (red solid curves) \cite{parameters}. Panel a) is for a 2D \textit{d}-wave superconductor, panel b) for a 3D Weyl superconductor (with $k_z=\pi/3$ at the Weyl point). The momentum follows a path through the magnetic Brillouin zone of Fig.\ \ref{fig_layout}. The location of the zero-field Dirac and Weyl points is indicated by green arrows. The $n$-th Landau level is expected at $E_n=\sqrt n\, E_1$, with $E_1=2\sqrt\pi\, v_{\rm F}/d_0$. In the \textit{d}-wave superconductor the Landau levels are destroyed by the vortex lattice \cite{Fra00}, while in the Weyl superconductor they are protected by chiral symmetry. 
}
\label{fig_dispersion}
\end{figure}

Here we propose that Weyl superconductors can make it happen. A Weyl semimetal with induced \textit{s}-wave superconductivity has massless nodal quasiparticles in a 3D Weyl cone \cite{Men12,Bed15}, with the same linear dispersion as the 2D Dirac cone of a \textit{d}-wave superconductor \cite{Vol93,Sim97}. We compare the band structures in Fig.\ \ref{fig_dispersion} \cite{appendix}. In zero magnetic field the gapless nodal points at the Fermi level ($E=0$) are qualitatively the same in both superconductors. But the response to a vortex lattice is fundamentally different: While in the \textit{d}-wave superconductor the dispersive Dirac cones persist, as expected \cite{Fra00}, in the Weyl superconductor a zeroth Landau level appears that is completely dispersionless in the plane perpendicular to the magnetic field. 

We will return to these numerical calculations later on, but first we want to explain why the zeroth Landau level in a Weyl superconductor is not broadened by the vortex lattice, as it is in a \textit{d}-wave superconductor. We have traced the origin of the difference to the topological protection of the zero-mode enforced by an index theorem for Hamiltonians with chiral symmetry \cite{Alv83}. For this explanation we will use an effective low-energy Hamiltonian. The numerics uses the full Hamiltonian and serves as a test of our analytics. We conclude with a discussion of the universal thermal conductance supported by the zero-mode.

\begin{figure}[tb]
\centerline{\includegraphics[width=0.9\linewidth]{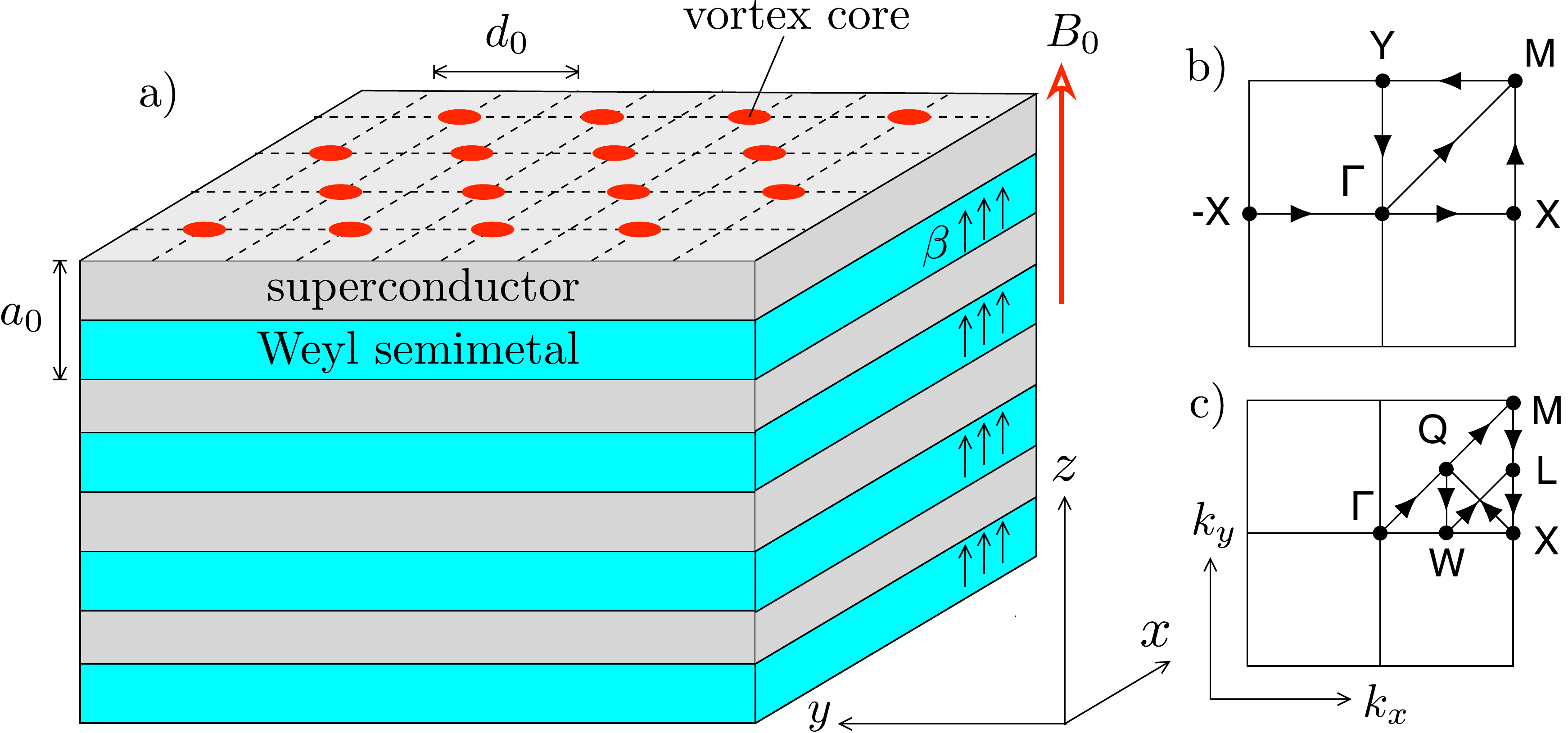}}
\caption{Weyl superconductor in the mixed phase. Panel a) shows a Weyl semimetal--superconductor heterostructure (layers of a topological insulator, with perpendicular magnetization $\beta$, separated by \textit{s}-wave superconducting spacer layers \cite{Men12}). A magnetic field $B_0$ is applied perpendicular to the layers. The heterostructure has lattice constant $a_0$, while the square vortex array has lattice constant $d_0$ (with two $h/2e$ vortices per unit cell). Panels b) and c) show two different paths through the magnetic Brillouin zone of the vortex array.
}
\label{fig_layout}
\end{figure}

\emph{Weyl superconductor in the mixed phase ---}
We start quite generally from the Bogoliubov-De Gennes (BdG) Hamiltonian in the Anderson gauge \cite{And98},
\begin{align}
&{\cal H}(\bm{k})=
U^\dagger\begin{pmatrix}
H_0(\bm{k}-e\bm{A})&\Delta\\
\Delta^\ast&-\sigma_y H_0^\ast(-\bm{k}-e\bm{A})\sigma_y
\end{pmatrix}
U\nonumber\\
&=
\begin{pmatrix}
H_0(\bm{k}+\bm{a}+m\bm{v}_s)&\Delta_0\\
\Delta_0&-\sigma_y H_0^\ast(-\bm{k}-\bm{a}+m\bm{v}_s)\sigma_y
\end{pmatrix},\label{HkAnderson}
\end{align}
with the definitions ($\hbar\equiv 1$, electron charge $+e$, mass $m$):
\begin{equation}
U=\begin{pmatrix}
e^{i\phi}&0\\
0&1
\end{pmatrix},\;\;\bm{a}=\tfrac{1}{2}\nabla\phi,\;\;m\bm{v}_s=\tfrac{1}{2}\nabla\phi-e\bm{A}.\label{Uavsdef}
\end{equation}
The $2\times 2$ matrix structure of $H$ refers to electron and hole quasiparticles, with single-particle Hamiltonian $H_0$ and its time-reverse in the diagonal blocks, coupled by the superconducting pair potential $\Delta=\Delta_0 e^{i\phi}$ in the off-diagonal blocks. The unitary transformation $U$ removes the spatially dependent phase $\phi(x,y)$ from the pair potential into the single-particle Hamiltonian, where it combines with the vector potential $\bm{A}(x,y)$ in the $x$--$y$ plane, corresponding to the magnetic field $\bm{B}=\nabla\times\bm{A}$ along $z$. 

Both the gauge field $\bm{a}(x,y)$ and the supercurrent velocity $\bm{v}_s(x,y)$ wind around the positions $\bm{R}_n$ of the vortex cores, according to
\begin{equation}
\nabla\times\nabla{\phi}=2\pi\hat{z}\textstyle{\sum_n}\delta(\bm{r}-\bm{R}_n).\label{curlgradphi}
\end{equation}
(For definiteness we assume the field points in the \textit{positive} $z$-direction.) A spatial average over the vortices gives a vanishing supercurrent velocity, $\overline{\bm{v}}_s=0$, while the average $\overline{\nabla\times\bm{a}}=e\bar{\bm{B}}$ gives the average magnetic field. The field is approximately uniform, equal to $B_0$, in the mixed phase $H_{c1}\ll B_0\ll H_{c2}$ of a type-II superconductor with overlapping vortices. In this regime the vortex cores occupy only a small fraction $B_0/H_{c2}\ll 1$ of the volume, so the amplitude $\Delta_0$ of the pair potential is also approximately uniform and only the phase $\phi$ is strongly position dependent.

We now specify to a Weyl superconductor, in the heterostructure configuration of Meng and Balents \cite{Men12,note5}: a stack in the $z$-direction of layers of Weyl semimetal alternating with an \textit{s}-wave superconductor. A magnetization $\beta$ perpendicular to the layers separates the Weyl cones in the Brillouin zone along $k_z$. The Weyl points are at $\bm{k}=(0,0,\pm K)$, $v_{\rm F}^2 K^2=\beta^2-\Delta_0^2$, with $v_{\rm F}$ the Fermi velocity (assumed isotropic for simplicity). The Weyl cones remain gapless as long as $\Delta_0<\beta$ \cite{note1}.

In the BdG Hamiltonian \eqref{HkAnderson} each Weyl cone is doubled into an electron and hole cone, mixed by the pair potential. We describe this mixing following Ref.\ \onlinecite{OBr17}, in the simplest case that the Weyl cones are close to the center $\bm{k}=0$ of the Brillouin zone and we may linearize the momenta. (All nonlinearities in the full Brillouin zone are included in our numerics.) The single-particle Weyl Hamiltonian $H_0$ is a $4\times 4$ matrix,
\begin{equation}
H_0(\bm{k})=v_{\rm F}\tau_z\bm{k}\cdot\bm{\sigma}+\beta\tau_0\sigma_z-\mu\tau_0\sigma_0,\label{H0linear}
\end{equation}
with $\mu$ the chemical potential. It is composed from Pauli matrices $\sigma_\alpha$ and $\tau_\alpha$ that act on the spin and orbital degree of freedom, respectively. We also need a third set of Pauli matrices $\nu_\alpha$ in the electron-hole basis. (The corresponding $2\times 2$ unit matrices are $\sigma_0$, $\tau_0$, $\nu_0$.) 

A unitary transformation ${\cal H}\mapsto V^\dagger {\cal H}V$ with
\begin{equation}
V=\exp(\tfrac{1}{2}i\theta\nu_y\tau_z\sigma_z),\;\;\tan\theta=-\frac{\Delta_0}{v_{\rm F}k_z},\;\;\theta\in(0,\pi),\label{Vdef} 
\end{equation}
followed by a projection onto the $\nu=\tau=\pm 1$ blocks, gives for the Weyl cones an effective $2\times 2$ low-energy Hamiltonian \cite{note2}:
\begin{align}
H_{\pm}(\bm{k})={}&v_{\rm F}\textstyle{\sum_{\alpha=x,y}}(k_\alpha+a_\alpha\pm \kappa mv_{s,\alpha})\sigma_\alpha\nonumber\\
&+(\beta-m_{k_z})\sigma_z\mp\kappa\mu\sigma_0,\label{HWeyldef}\\
m_{k_z}={}&\sqrt{\Delta_0^2+v_{\rm F}^2 k_z^2},\;\;\kappa=-v_{\rm F}k_z/m_{k_z}.\label{kappadef}
\end{align}
The electron-like and hole-like cones have opposite sign of the effective charge $q_{\rm eff}=\pm\kappa e$, with $|q_{\rm eff}|\rightarrow e\sqrt{1-\Delta_0^2/\beta^2}$ for $|k_z|\rightarrow K$, smaller than the bare charge $e$ due to the mixing of electrons and holes by the pair potential \cite{Bai17}. The velocity $v_z=\partial m_{k_z}/\partial k_z$ perpendicular to the layers is also renormalized by the superconductivity: $v_z\rightarrow v_{\rm F}^2 K/\beta$ for $|k_z|\rightarrow K$.

At the Weyl point, for $\mu=0$ and $|k_z|=K$, the Hamiltonian \eqref{HWeyldef} anticommutes with $\sigma_z$. This socalled chiral symmetry gives a formal correspondence with a problem first studied 40 years ago by Aharonov and Casher \cite{Aha79}, as an application of an index theorem from supersymmetric quantum mechanics \cite{Alv83}. The problem of Ref.\ \onlinecite{Aha79}, to determine the zeroth Landau level of a two-dimensional massless electron in an inhomogeneous magnetic field, has also been studied more recently in the context of graphene \cite{Kat08,Kai09,Lu14}. We need to adapt the calculation here to account for the fractionally charged quasiparticles, but the basic approach carries through.

\emph{Calculation of the zero-modes ---}
To study the effect of chiral symmetry on the Landau level spectrum we set $\mu=0$, $|k_z|=K$ and focus our attention on the chiral Hamiltonian
\begin{equation}
\begin{split}
&H_{\rm chiral}=v_{\rm F}\begin{pmatrix}
0&{\cal D}\\
{\cal D}^\dagger&0
\end{pmatrix},\;\;{\cal D}=\Pi_x-i\Pi_y,\\
&\bm{\Pi}=-i\nabla+e{\cal A},\;\;
e{\cal A}=\bm{a}\pm\kappa m\bm{v}_s.
\end{split}
\label{Hchiraldef}
\end{equation}
(We omit the $\pm$ subscript for ease of notation.) The effective vector potential ${\cal A}$ describes the effective magnetic field
\begin{equation}
{\cal B}=\partial_x {\cal A}_y-\partial_y {\cal A}_x=\Phi_0(1\pm\kappa)\textstyle{\sum_n}\delta(\bm{r}-\bm{R}_n)\mp\kappa B
\end{equation}
felt by the Weyl fermions in the vortex lattice. 

For what follows it is convenient to choose a gauge such that $\nabla\cdot{\cal A}=0$ and to assume that the external magnetic field $B_0$ is imposed on a large but finite area $S$. Because there are $N_{\text{vortex}}=B_0 S/\Phi_0$ vortices in that area (with $\Phi_0=h/2e$ the superconducting flux quantum), the flux $\Phi=\int d\bm{r}\,{\cal B}= B_0 S$ through the system corresponding to the effective field equals the real flux. (The $\kappa$-dependence of ${\cal B}$ drops out upon spatial integration.)

A zero-mode $\psi$ of $H_{\rm chiral}$ is either a spinor $u\choose 0$ with ${\cal D}^\dagger u=0$ or it is a spinor $0\choose v$ with ${\cal D}v=0$. The general solution of these two differential equations has the form \cite{Aha79,Kai09,note3}:
\begin{equation}
\begin{split}
&u=f(\zeta)e^W,\;\;v=f(\zeta^\ast)e^{-W},\;\;\zeta=x+iy,\\
&W(\bm{r})=\frac{1}{2\Phi_0}\int dx'\int dy'\,{\cal B}(\bm{r}')\ln|\bm{r}-\bm{r}'|.
\end{split}
\label{uvWdef}
\end{equation}
The difference ${\cal N}=N_u-N_v$ in the number of normalizable solutions for $u$ and $v$ is called the index of $H_{\rm chiral}$. The absolute value $|{\cal N}|$ is a lower bound on the degeneracy of the zero-mode and the sign of $\cal N$ determines the chirality: whether the zero-mode is an eigenstate of $\sigma_z$ with eigenvalue $+1$ or $-1$.

To determine the index of $H_{\rm chiral}$ we proceed as follows. In the absence of vortices the function $f(\zeta)$ is analytic in the entire complex plane and we can use a basis of polynomials. A polynomial $f(\zeta)$ of degree $N-1$ then produces $N$ linearly independent zero-modes --- provided $u$ or $v$ is normalizable, $\int rdr\,|\psi|^2<\infty$. For large $r$ one has asymptotically
\begin{equation}
W\rightarrow \tfrac{1}{2}(\Phi/\Phi_0)\ln |\bm{r}|\Rightarrow e^{W}\rightarrow |\bm{r}|^{ N_{\text{vortex}}/2},\label{largerasymp}
\end{equation}
so if only the decay at infinity would be an issue we would conclude that $N_u=0$, $N_v=\text{Int}\,[N_{\text{vortex}}/2]$. This is the answer in the absence of vortices \cite{Aha79}, when the degeneracy of the zero-mode is determined by the enclosed flux in units of $h/e=2\Phi_0$, while the chirality is set by the sign of the magnetic field (which we have assumed positive). As we will now show, the presence of vortices introduces a dependence of the chirality on the sign of the fractional charge $q_{\rm eff}=\pm\kappa e$ of the quasiparticles, while the degeneracy remains given by the bare electron charge $e$.

With vortices the function $f(\zeta)$ may have poles at the vortex cores $\zeta_n=x_n+iy_n$. We use this freedom to re-express the solution \eqref{uvWdef} as
\begin{equation}
u=g(\zeta)e^W\textstyle{\prod_n}(\zeta-\zeta_n)^{-1},\;\;v=f(\zeta^\ast)e^{-W}.\label{uvWdefnew}
\end{equation}
If for $f$ and $g$ we take polynomials of degree $N-1$, with $N=\text{Int}\,[N_{\text{vortex}}/2]$, then both the functions $u$ and $v$ decay sufficiently rapidly at infinity. The boundary condition at the vortex cores now determines which of the two solutions is realized.

Near a vortex at position $\bm{r}_n$ the asymptotics is
\begin{equation}
|u|^2\rightarrow|\bm{r}-\bm{r}_n|^{-1+q_{\rm eff}/e},\;\;|v|^2\rightarrow|\bm{r}-\bm{r}_n|^{-1-q_{\rm eff}/e}.\label{smallrasymp}
\end{equation}
Since $|q_{\rm eff}|<e$ both solutions $\psi_u={u\choose 0}$ and $\psi_v={0\choose v}$ remain square integrable at the vortex core. The boundary condition \cite{appendix}
\begin{equation}
\sigma_z\psi=(\text{sign}\,q_{\rm eff})\,\psi,\;\;\text{for}\;\;\bm{r}\rightarrow\bm{r}_n.\label{psiboundary}
\end{equation}
selects the most weakly divergent solution in Eq.\ \eqref{smallrasymp}: $\psi=\psi_u$ with positive chirality for $q_{\rm eff}>0$ and $\psi=\psi_v$ with negative chirality for $q_{\rm eff}<0$.

All of this was for $\mu=0$, $|k_z|=K$, but both terms $\mu\sigma_0$ and $(\beta-m_{k_z})\sigma_z$ from Eq.\ \eqref{HWeyldef} can be immediately reinstated since the zero-mode is an eigenstate of $\sigma_z$. The resulting $\mu$ and $k_z$-dependence of the zeroth Landau level is
\begin{equation}
E_\pm(k_z)=\mp\kappa\mu+({\rm sign}\,q_{\rm eff})(\beta-m_{k_z}).\label{EpmzerothLL}
\end{equation}

We have thus seen how the chiral symmetry protects the zeroth Landau level from being destroyed by the vortex lattice. To complete this analytical treatment, we point out why the \textit{d}-wave superconductor lacks a similar protection. In the Anderson gauge, the low-energy Hamiltonian near the nodal point of a \textit{d}-wave pair potential reads \cite{And98,Fra00,Vaf01a}
\begin{equation}
H_{d\text{-wave}}=v_{\rm F}(k_x+a_x)\sigma_z+v_\Delta(k_y+a_y)\sigma_x+mv_{s,x}\sigma_0.\label{Hdwave}
\end{equation}
There are inessential differences with $H_{\rm chiral}$ from Eq.\ \eqref{Hchiraldef} --- the Dirac cone is anisotropic and the basis of Pauli matrices is rotated --- but the essential difference is that the superfluid velocity breaks the chiral symmetry: $H_{d\text{-wave}}$ anticommutes with $\sigma_y$ only if $v_{s,x}=0$. In the \textit{d}-wave superconductor the superfluid velocity enters as a chirality-breaking scalar potential, while in the Weyl superconductor it is a chirality-preserving vector potential. The former is a strong scatterer, which effectively destroys the Landau levels, while the latter cannot by force of the topological index theorem.

\begin{figure}[tb]
\centerline{\includegraphics[width=0.8\linewidth]{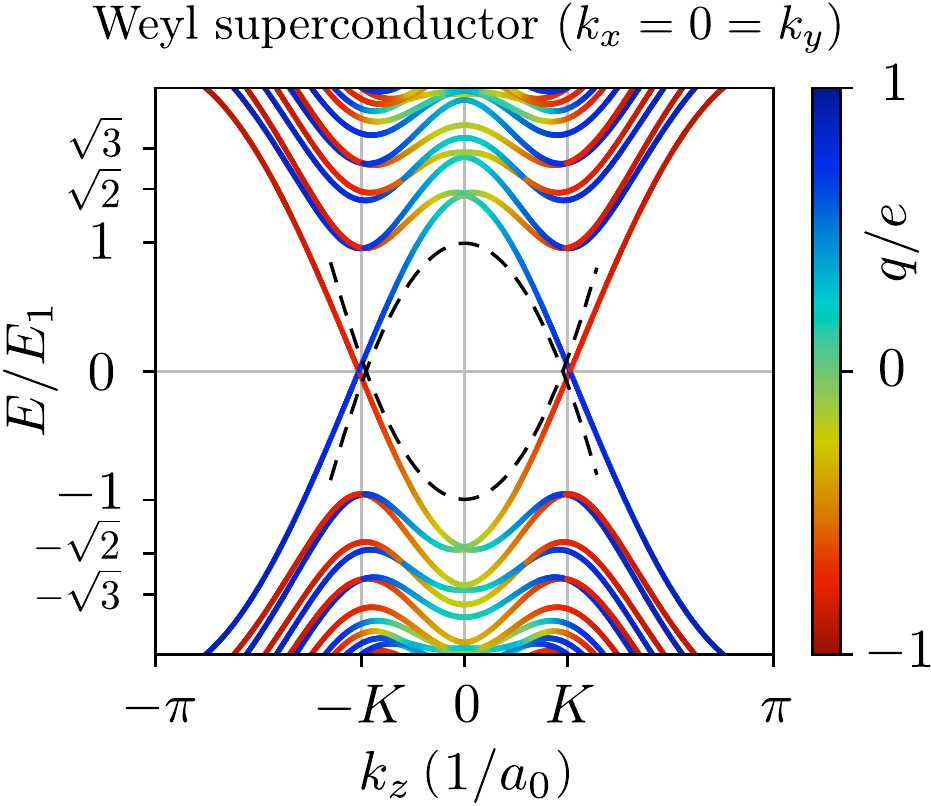}}
\caption{Same as Fig.\ \ref{fig_dispersion}b, but now as a function of $k_z$ for $k_x=0=k_y$ at the center of the Brillouin zone \cite{param2}. The color scale indicates the charge expectation value. The dashed curve is the dispersion \eqref{EpmzerothLL} of the zeroth Landau level, calculated analytically for $K\ll 1$ (which explains the deviation from the numerics). The effective charge at $E=0$ is $\pm 0.73$, close to the analytical prediction of $\pm\kappa=\pm 1/\sqrt 2$.
 }
\label{fig_kzdispersion}
\end{figure}

\begin{figure}[tb]
\centerline{\includegraphics[width=0.8\linewidth]{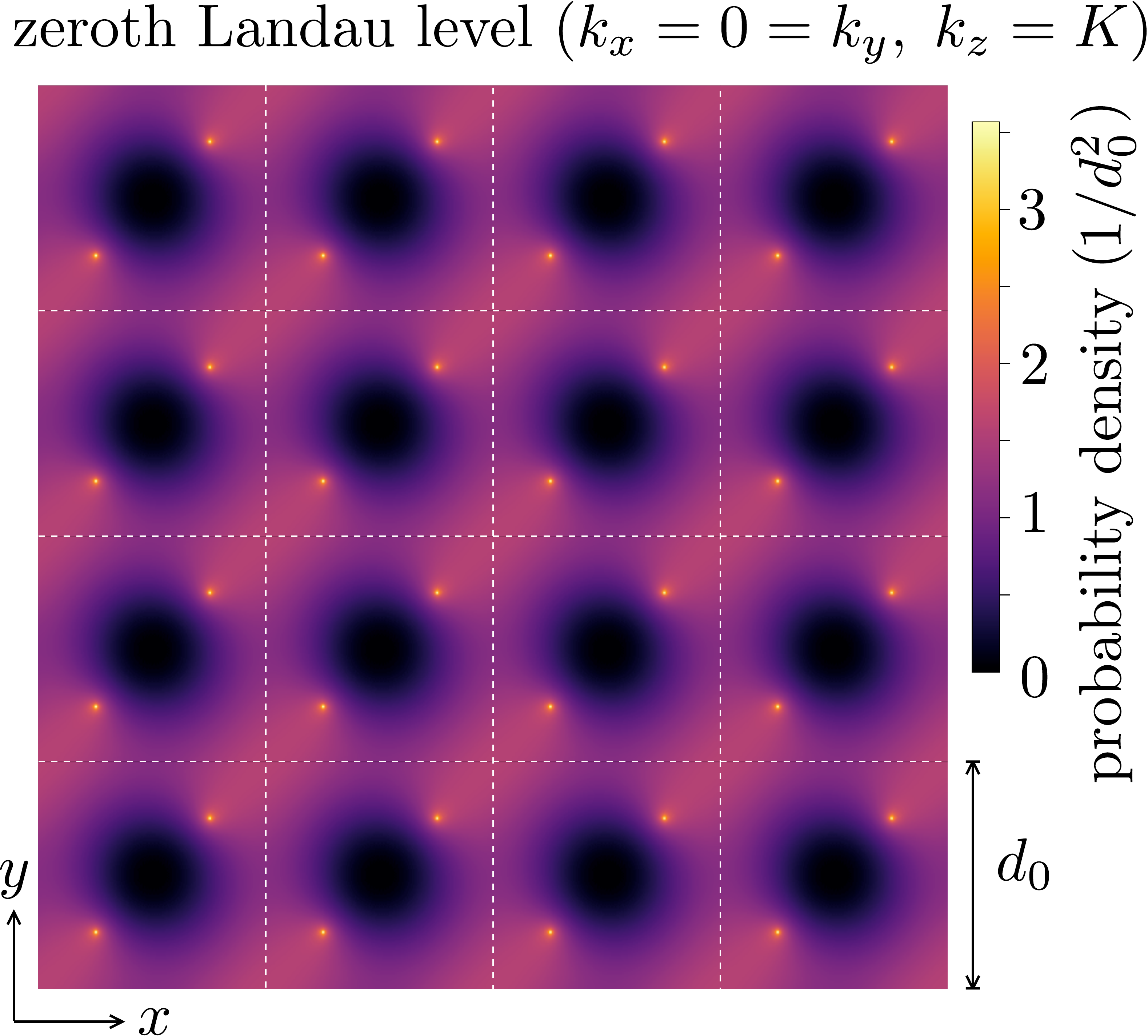}}
 \caption{Color scale plot of $|\psi(x,y)|^2$ in the zeroth Landau level of the Weyl superconductor \cite{param3}. The white dashed lines indicate the vortex array, with a pair of $h/2e$ vortices in each unit cell. On approaching a vortex core, when the separation $\delta r\rightarrow 0$, the density diverges as a power law $|\psi|^2\propto \delta r^{1/\sqrt 2-1}$, in accord with Eq.\ \eqref{smallrasymp}. 
}
\label{fig_vortexarray}
\end{figure}

\emph{Comparison with numerics ---}
To test our analytical theory we have numerically calculated the spectrum of a Weyl superconductor with a vortex lattice, using the \textit{Kwant} tight-binding code \cite{kwant}. The $8\times 8$ Hamiltonian has the BdG form \eqref{HkAnderson} with \cite{Men12,Bed15,Bov17}
\begin{align}
H_0(\bm{k})={}&t_0{\sum_{\alpha=x,y,z}}\left[\tau_z\sigma_\alpha\sin k_\alpha a_0+\tau_x\sigma_0(1-\cos k_\alpha a_0)\right]\nonumber\\
&+\beta\tau_0\sigma_z-\mu\tau_0\sigma_0.\label{H0Weyl}
\end{align}
Near the center of the Brillouin zone this reduces to the linearized Hamiltonian \eqref{H0linear}, but now we will not make any linearization. Results are shown in Figs.\ \ref{fig_dispersion}b, \ref{fig_kzdispersion}, and \ref{fig_vortexarray} \cite{appendix}. They are fully consistent with the analytics.

\emph{Thermal conductance ---}
The chiral zeroth Landau level governs the thermal transport properties of the Weyl superconductor, in the direction parallel to the magnetic field. The degeneracy $eB_0S/h=\tfrac{1}{2}\Phi/\Phi_0$ of the zeroth Landau level implies a thermal conductance
\begin{equation}
G_=\tfrac{1}{2}g_0\Phi/\Phi_0,\;\;g_0={\cal L}Te^2/h,\label{Gthermal}
\end{equation}
with ${\cal L}=\tfrac{1}{3}(\pi k_{\rm B}/e)^2$ the Lorenz number. In words, each vortex contributes half a thermal conductance quantum to the heat transport --- the factor $1/2$ being a reminder that the quasiparticles in the Weyl superconductor are Majorana fermions \cite{Bai17}. Do note that the states in the zeroth Landau level are extended over the $x$--$y$ plane, the current flow is not confined to the vortex cores (see Fig.\ \ref{fig_vortexarray}) \cite{note4}. We expect the universal thermal conductance \eqref{Gthermal} to be robust against non-magnetic disorder, which in the effective Hamiltonian would enter as a term $\propto\sigma_z$ that does not couple Landau levels of opposite chirality.

\emph{Conclusion ---}
In this work we have revisited the celebrated question \cite{Gor98,And98} whether quasiparticles in the vortex lattice of a gapless superconductor can condense into Landau levels. We have shown that Weyl superconductors can accomplish what \textit{d}-wave superconductors could not \cite{Fra00}: The chirality of Weyl fermions protects the zeroth Landau level from broadening due to scattering by the vortices. We have developed the analytical argument for a simple low-energy Hamiltonian and supported it by numerical calculations for a heterostructure model of the Weyl superconductor \cite{Men12}. 
We anticipate that the Landau levels will govern the thermodynamic and transport properties of the vortex lattice, finally allowing for the observation of quantum effects that proved elusive in the \textit{d}-wave context. 

\emph{Acknowledgements ---}
We have benefited from discussions with D. I. Pikulin and J. Tworzyd{\l}o. This research was supported by the Netherlands Organization for Scientific Research (NWO/OCW), an ERC Synergy Grant, and by the T\"{U}B\.{I}TAK grant No.\ 114F163.

\clearpage


\appendix

\makeatletter
\renewcommand{\bibnumfmt}[1]{[S#1]}
\renewcommand{\citenumfont}[1]{S#1}

\section{Boundary condition at the vortex core}
\label{App_bc}

We consider the chiral Hamiltonian \eqref{Hchiraldef} near a vortex at the origin,
\begin{equation}
H_{\rm vortex}=v_{\rm F}\sum_{\alpha=x,y}(p_\alpha+e{\cal A}_\alpha)\sigma_\alpha+M(\bm{r})\sigma_z,\label{Hvortex}
\end{equation}
retaining only the singular contribution to the vector potential,
\begin{equation}
\nabla\times e{\cal A}=(e+q_{\rm eff})\Phi_0\hat{z}\delta(\bm{r})\Rightarrow e{\cal A}=\frac{(e+q_{\rm eff})\Phi_0}{2\pi r}\hat{\theta}.\label{Avortex}
\end{equation}
A similar eigenvalue problem has been studied in the context of graphene \cite{DeM10}, but without the fractional charge $q_{\rm eff}=\pm \kappa e$ characteristic of the Weyl superconductor.

We model the delta-function vortex singularity by a mass term $M(\bm{r})=M_0\theta(d_{\rm vortex}-r)$, in the limit $M_0\rightarrow\infty$, $d_{\rm vortex}\rightarrow 0$ with $M_0 d_{\rm vortex}^2$ finite. In that limit the effective charge tends to the bare charge, $q_{\rm eff}\rightarrow \pm e$, within the vortex core.

In polar coordinates $(r,\theta)$ one has
\begin{subequations}
\label{polarcoord}
\begin{align}
&\frac{\partial}{\partial x}+i\frac{\partial}{\partial y}=e^{i\theta}\left(\frac{\partial}{\partial r}+\frac{i}{r}\frac{\partial}{\partial \theta}\right),\label{polarcoorda}\\
&e{\cal A}_x+ie{\cal A}_y=\frac{\lambda}{r}ie^{i\theta},\;\;\lambda=\tfrac{1}{2}+q_{\rm eff}/2e\in(0,1).\label{polarcoordb}
\end{align}
\end{subequations}
(Recall that $e\Phi_0/2\pi=\hbar/2\equiv 1/2$.) The Dirac Hamiltonian then takes the form
\begin{subequations}
\label{Hvortex2}
\begin{align}
&H_{\rm vortex}=\begin{pmatrix}
M&D_-\\
D_+&-M
\end{pmatrix},\label{Hvortex2a}\\
&D_\pm=v_{\rm F}e^{\pm i\theta}\left(-i\frac{\partial}{\partial r}\pm\frac{1}{r}\frac{\partial}{\partial \theta}\pm\frac{i\lambda}{r}\right).
\label{Hvortex2b}
\end{align}
\end{subequations}

Since $H_{\rm vortex}$ commutes with the angular momentum operator $J=-i\partial_\theta+\tfrac{1}{2}\sigma_z$, with eigenvalues $m-1/2$ for integer $m$, the eigenstates of $H_{\rm vortex}$ can be chosen as eigenstates of $J$,
\begin{subequations}
\label{psimdef}
\begin{align}
&\psi_m(r,\theta)=e^{im\theta}\begin{pmatrix}
e^{-i\theta}u_m(r)\\
iv_m(r)
\end{pmatrix},\label{psimdefa}\\
&(M-E)u_m+v_{\rm F}[\partial_r+(m+\lambda)r^{-1}]v_m=0,\label{psimdefb}\\
&(M+E)v_m+v_{\rm F}[\partial_r-(m-1+\lambda)r^{-1}]u_m=0.\label{psimdefc}
\end{align}
\end{subequations}

We take $E=0$ and consider the solutions outside the vortex core ($r>d_{\rm vortex}$, where $M=0$) and inside the vortex core ($r<d_{\rm vortex}$, $M=M_0>0$). Outside the vortex core the solutions for $u_m$ and $v_m$ decouple,
\begin{equation}
u_m=C_1r^{m-1+\lambda},\;\;v_m=C_2r^{-m-\lambda},\label{umvmoutside}
\end{equation}
with independent coefficients $C_1,C_2$. Inside the vortex core we have, in view of the Bessel function identities
\begin{subequations}
\label{Besselidentity}
\begin{align}
&\partial_r I_\alpha(r)\pm(\alpha/r) I_\alpha(r)=I_{\alpha\mp 1}(r),\label{Besselidentitya}\\
&\partial_r K_\alpha(r)\pm(\alpha/r) K_\alpha(r)=-K_{\alpha\mp 1}(r),\label{Besselidentityb}
\end{align}
\end{subequations}
the general solution
\begin{equation}
\begin{split}
&u_m(r)=C_3\,I_{m-1+\lambda}(M_0\, r/v_{\rm F})+C_4 K_{m-1+\lambda}(M_0\, r/v_{\rm F}),\\
&v_m(r)=-C_3\,I_{m+\lambda}(M_0\,r/v_{\rm F})+C_4 K_{m+\lambda}(M_0\,r/v_{\rm F}).
\end{split}\label{umvminside}
\end{equation}
We may set $C_4=0$ to obtain a regular solution at $r=0$ for $q_{\rm eff}=\pm e\Rightarrow \lambda\in\{0,1\}$.

The global solution \eqref{uvWdefnew} has outside the vortex at $\bm{r}_n\equiv 0$ the asymptotics
\begin{equation}
\psi_{\rm outside}=\begin{pmatrix}
C_1e^{-i\theta}r^{-1/2+q_{\rm eff}/2e}\\
C_2 ir^{-1/2-q_{\rm eff}/2e}
\end{pmatrix},\label{psi0outside}
\end{equation}
since $\zeta-\zeta_n=e^{i\theta}r$. This corresponds to the local solution $\psi_{m}(r,\theta)$ outside the vortex core for quantum number $m=0$. We need to match this to the $m=0$ solution inside the vortex core. In the large-$M_0$ limit, for $M_0 \gg v_{\rm F}/r$, this has the asymptotics 
\begin{equation}
\psi_{\rm inside}=\frac{C_3\, e^{M_0 r/v_{\rm F}}}{\sqrt{2\pi M_0r/v_{\rm F}}}\begin{pmatrix}
e^{-i\theta}\\
-i
\end{pmatrix},\label{psi0inside}
\end{equation}
since the Bessel-$K$ function becomes exponentially small $\propto\exp(-M_0 r/v_{\rm F})$.

Equating $\psi_{\rm outside}$ and $\psi_{\rm inside}$ at $r=d_{\rm vortex}$ gives the ratio of coefficients
\begin{equation}
C_2/C_1=-(d_{\rm vortex})^{q_{\rm eff}/e}.\label{C2C2ratio}
\end{equation}
If we finally send $d_{\rm vortex}\rightarrow 0$, we find that $C_2\rightarrow 0$ for $q_{\rm eff}>0$, while $C_1\rightarrow 0$ for $q_{\rm eff}<0$. This corresponds to the boundary condition \eqref{psiboundary} in the main text.

\section{Details of the tight-binding calculations}
\label{tightbinding}

\subsection{Weyl superconductor}

We discretize the BdG Hamiltonian \eqref{HkAnderson} in the Anderson gauge on a square lattice, lattice constant $a_0\equiv 1$, nearest-neigbor hopping energy $t_0\equiv 1$. For the diagonal block $H_0(\bm{k})$ we take the four-band model of Eq.\ \eqref{H0Weyl}. The tight-binding Hamiltonian is
\begin{widetext}
\begin{align}
    {\cal H} ={}& 
    \sum_{\bm n} \begin{pmatrix}
        h(k_z) & \Delta_0 \\
        \Delta_0 & -\sigma_y h(-k_z)^* \sigma_y
\end{pmatrix}
    |\bm n\rangle\langle \bm n|\nonumber\\
    & + 
   \frac{1}{2} \sum_{\bm n,\hat{\bm\delta}}
    \begin{pmatrix}
\exp\left({i}\int_{\bm n}^{\bm{n}+\hat{\bm\delta}} e\bm A\cdot{d} \bm l - {i}\phi_{\bm n+\hat{\bm\delta}} + {i}\phi_{\bm n}\right) &0 \\ 
0& -\exp\left(-{i}\int_{\bm n}^{\bm n+\hat{\bm\delta}} e\bm A\cdot{d} \bm l\right)
\end{pmatrix}        
({i} \tau_z\bm\sigma\cdot{\hat{\bm\delta}} -   \tau_x\sigma_0) |\bm n+\hat{\bm\delta}\rangle\langle \bm n|,\\
    h(k_z) &= \tau_z\sigma_z  \sin k_z+ 
    \tau_x \sigma_0 (3-\cos k_z)
    + \beta\tau_0\sigma_z  - \mu\tau_0\sigma_0.
\end{align}
\end{widetext}
The vector $\bm{n}$ labels the lattice sites and the unit vector $\hat{\bm \delta}$ points to the four nearest neighbors. We denote by $\phi_{\bm{n}}$ the superconducting phase $\phi(\bm{r})$ at site $\bm{n}$.

We assume a uniform magnetic field $\bm{B}=B_0\hat{z}$ (appropriate for the strong-type-II regime $H_{c1}\ll B_0\ll H_{c2}$), with vector potential
\begin{equation}
\bm A(x,y) =-\frac{2\pi}{eN^{2}} (y,0,0)
\end{equation}
corresponding to a flux $h/e$ through a supercell of $N\times N$ unit cells (square magnetic unit cell, lattice constant $d_0=Na_0$). The conjugate vector potential
\begin{equation}
\bar{\bm{A}}(x,y)=-\frac{2\pi}{eN^{2}} (0,x,0)
\end{equation}
is defined such that $\bm{\Pi}=\bm{p}-e\bm{A}$ and $\bar{\bm \Pi}=\bm{p}-e\bar{\bm{A}}$ commute, $[\Pi_\alpha,\Pi_\beta]=0$. It enters in the magnetic periodic boundary conditions \cite{Bro64,Zak64,Fis70}
\begin{equation}
\begin{split}
    &\psi(N,y) = e^{iN[k_x-e\bar{A}_x(0,y)] }\psi(0,y)=e^{{i} k_x N} \psi(0,y),\\
    & \psi(x,N) = e^{iN[k_y-e\bar{A}_y(x,0)] }\psi(x,0)=e^{{i} k_y N +  2\pi ix/N}\psi(x,0),
\end{split}
\end{equation}
for $x,y\in\{0,1,\dots,N-1\}$. 

\begin{figure}[tb]
\centerline{\includegraphics[width=1\linewidth]{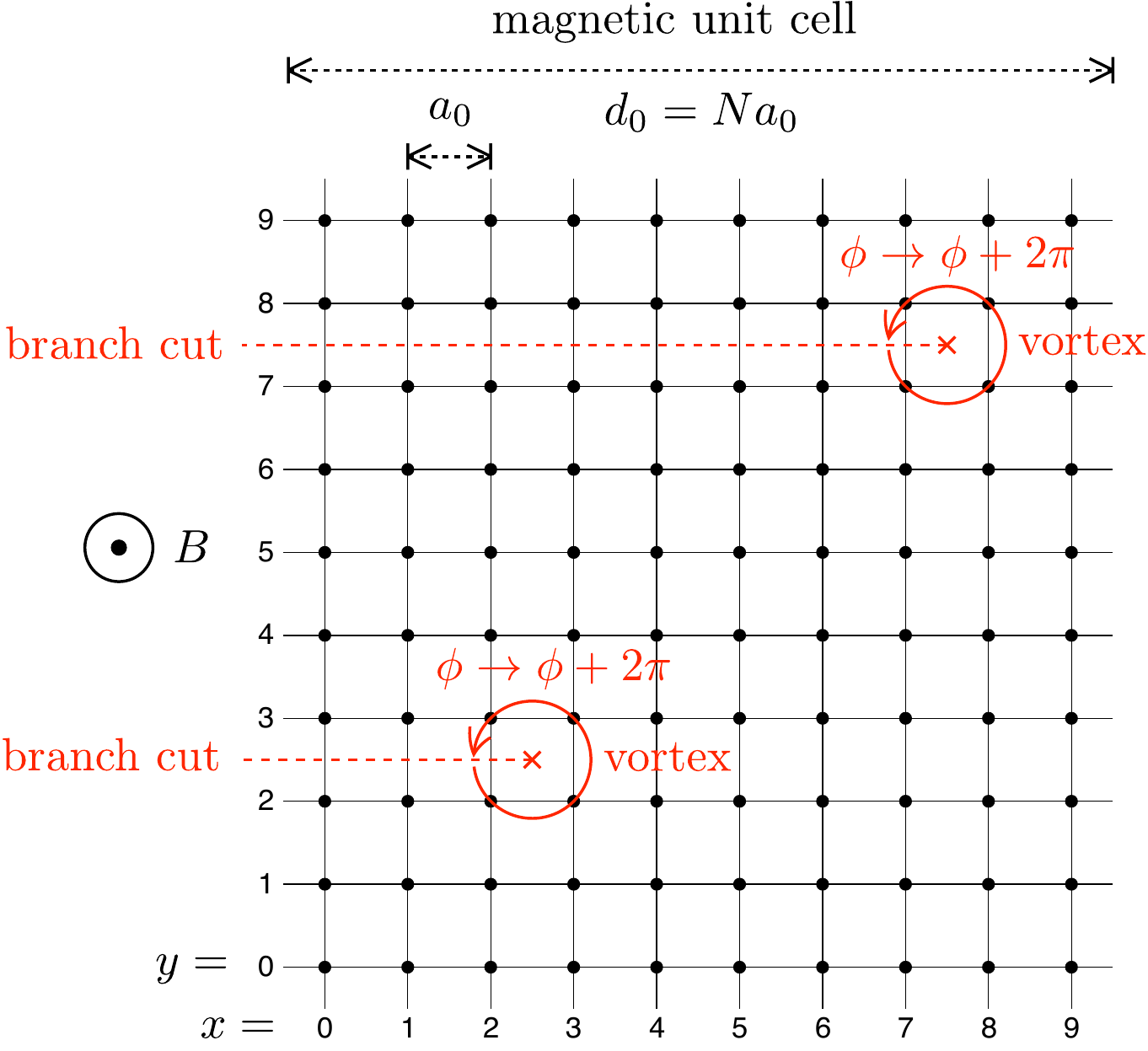}}
\caption{Magnetic unit cell for $N=10$, containing a pair of $h/2e$ vortices at the positions specified by Eq.\ \eqref{xyvortex}. The superconducting phase winds by $2\pi$ upon encircling a vortex, producing a branch cut. At the two sides $(x,y\pm\epsilon)$ of a branch cut one has $\phi(x,y+\epsilon)=\phi(x,y-\epsilon)+2\pi$ . }
\label{fig:diagram}
\end{figure}

In each supercell we place a pair of $h/2e$ vortices, at positions
\begin{equation}
\begin{split}
&    x_{\rm vortex}^{(1)} = y_{\rm vortex}^{(1)} = {\rm Int}\,[ N/4 ] + 1/2,\\
&    x_{\rm vortex}^{(2)} = y_{\rm vortex}^{(2)} = N-1/2-{\rm Int}\,[ N/4 ],
\end{split}\label{xyvortex}
\end{equation}
see Fig.\ \ref{fig:diagram}. This produces a square vortex array consisting of two sublattices with lattice constant $d_0$.

\begin{figure}[tb]
\centerline{\includegraphics[width=0.7\linewidth]{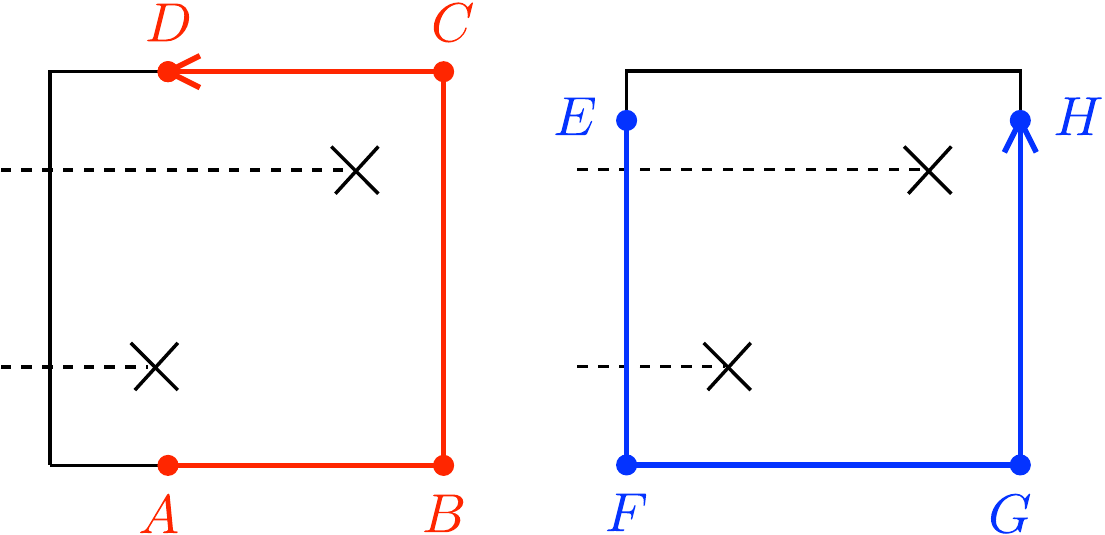}}
\caption{Two integration paths $C$ along the boundary of the magnetic unit cell for which $\int_C\bm{v}_s\cdot d\bm{l}=0$, as a consequence of Eq.\ \eqref{vssymmetry}. Vortices are indicated by crosses, the branch cuts in the phase by dashed lines. For the red path the integral along segment $BC$ vanishes, while the contributions from the segments $AB$ and $CD$ cancel. For the blue path the segment $FG$ does not contribute and $EF$ cancels with $GH$.}
\label{fig:conts}
\end{figure}

\subsection{Superconducting phase}

In the continuum description the phase $\phi(\bm{r})$ of the superconducting order parameter is determined by
\begin{equation}
\nabla\times\nabla\phi=\sum_n\delta(\bm{r}-\bm{r}_n),\;\;\nabla\cdot\nabla\phi=0.\label{phiequations}
\end{equation}
The first equation specifies a $2\pi$ winding of the phase around each vortex, at position $\bm{r}_n$, and the second equation ensures that the supercurrent velocity $m\bm{v}_s=\tfrac{1}{2}\nabla\phi-e\bm{A}$ has vanishing divergence. (Note that $\nabla\cdot\bm{A}=0$ for our choice of gauge.) 

We discretize Eq.\ \eqref{phiequations} in the $N\times N$ magnetic unit cell of Fig.\ \ref{fig:diagram}. To each of the two vortices in this supercell we assign a branch cut running from $(x_{\rm vortex},y_{\rm vortex})$ to $(0,y_{\rm vortex})$, at which the phase jumps by $2\pi$. The discrete version of Eq.\ \eqref{phiequations} then reads
\begin{align}
   &  \phi(x,y-1)+ \phi(x+1,y) + \phi(x-1,y) + \phi(x,y+1)\nonumber\\
  \qquad & -4\phi(x,y)   = \begin{cases}
 \pm 2\pi & \text{if $(x,y)\rightarrow(x,y\pm 1)$}\\
  & \text{crosses a branch cut,} \\
        0 & \text{otherwise,}
\end{cases}      \label{phiLaplace}
\end{align}
for $x,y\in\{0,1,2,\ldots N-1\}$.

We need to supplement Eq.\ \eqref{phiLaplace} by periodic boundary conditions at the edges of the magnetic unit cell. To determine these we integrate
\begin{equation}
    \phi(\bm{r}) - \phi(\bm{r}') =2\int_{\bm{r}'}^{\bm{r}} (m\bm{v}_s+e\bm{A})\cdot{d}\bm{l} + 2\pi n\label{Cintegrate}
\end{equation}
along a path $C$ from $\bm{r}'$ to $\bm{r}$. The discontinuity of $\phi$ when $C$ crosses a branch cut is accounted for by the $2\pi n$ offset: The integer $n$ equals the number of branch cut lines crossed from below minus those crossed from above. 

The trick is to choose a path such that the integral of the supercurrent velocity vanishes. The combination of periodicity and inversion symmetry implies that
\begin{equation}
\begin{split}
\bm{v}_s(x,y)={}&\bm{v}_s(x+N,y)=\bm{v}_s(x,y+N),\\
\bm{v}_s(x,y)={}&-\bm{v}_s(-x,-y)\\
\Rightarrow{}&\bm{v}_s(N,y)=-\bm{v}_s(N,N-y),\\
&\bm{v}_s(x,0)=-\bm{v}_s(N-x,0).
\end{split}\label{vssymmetry}
\end{equation}
As a consequence, the integral $\int_C\bm{v}_s\cdot d\bm{l}=0$ vanishes for the two paths of Fig.\ \ref{fig:conts}. Integration of the vector potential gives the boundary conditions.
\begin{subequations}
\label{phibc}
\begin{align}
    \phi(x,y_0+N) &= \phi(x,y_0) + 4\pi(1-x/N)\,,\\
    \phi(x_0+N,y) &= \phi(x_0,y) - 2\pi\times(\text{number of branch cuts}\nonumber\\
    	&\qquad\qquad\qquad\qquad\qquad\text{ below $y$})\,,
\end{align}
\end{subequations}
where $x_0,y_0\in\{0,-1\}$.

The set of equations \eqref{phiLaplace} and \eqref{phibc} can be written in a matrix form, $ \sum_j M_{ij}\phi_j = b_i$ for a real symmetric matrix $M$, which we solved using the conjugate gradient method.

\begin{widetext}

\subsection{$\bm{d}$-wave superconductor}

A 2D superconductor with spin-singlet $d_{x^2-y^2}$ pairing symmetry has BdG Hamiltonian
\begin{equation}
\begin{split}
&{\cal H} = \begin{pmatrix}
\frac{1}{2m}(\bm{k}-e\bm{A})^2 -\mu& (\bm{k}-e\bm{A})\cdot\bm{\Delta}\cdot(\bm{k}+e\bm{A}) \\
(\bm{k}+e\bm{A})\cdot\bm{\Delta}^\dagger\cdot(\bm{k}-e\bm{A}) & -\frac{1}{2m}(\bm{k}+e\bm{A})-\mu
\end{pmatrix},\\
&\bm{k}=(k_x,k_y)=-i\hbar(\partial_x,\partial_y),\;\;\bm{\Delta}(\bm{r})=\Delta_0 e^{i\phi(\bm{r})}\begin{pmatrix}
1&0\\
0&-1
\end{pmatrix}.
\end{split}
\label{Hdwavedef}
\end{equation}
Our choice of symmetrization of the pair potential follows Ref.\ \onlinecite{Die13}. One checks that the choice \eqref{Hdwavedef} satisfies the requirement of gauge invariance,
\begin{equation}
\begin{pmatrix}
e^{-i\chi}&0\\
0&e^{i\chi}
\end{pmatrix}{\cal H}(e\bm{A},\bm{\Delta})\begin{pmatrix}
e^{i\chi}&0\\
0&e^{-i\chi}
\end{pmatrix}={\cal H}(e\bm{A}-\nabla\chi,e^{-2i\chi}\bm{\Delta}).\label{gaugeinvariance}
\end{equation}

Following Ref.\ \onlinecite{Die13} we discretize ${\cal H}$ on a square lattice (lattice constant $a_0\equiv 1$, nearest neighbor hopping energy $t_0=\hbar^2/2ma_0^{2}$). At the end we carry out the Anderson gauge transformation,
\begin{equation}
{\cal H}\mapsto\begin{pmatrix}
e^{-i\phi}&0\\
0&1
\end{pmatrix}{\cal H}\begin{pmatrix}
e^{i\phi}&0\\
0&1
\end{pmatrix}.
\end{equation}
The resulting tight-binding Hamiltonian
\begin{equation}
{\cal H}=\sum_{\bm{m},\bm{n}}\begin{pmatrix}
t_{ee}(\bm{m},\bm{n})&t_{eh}(\bm{m},\bm{n})\\
t_{he}(\bm{m},\bm{n})&t_{hh}(\bm{m},\bm{n})
\end{pmatrix}|\bm{m}\rangle\langle\bm{n}|
\end{equation}
has nonzero matrix elements for $\bm{m}=\bm{n}$ and $\bm{m}=\bm{n}+\hat{\bm{\delta}}$, with $\hat{\bm{\delta}}\in\{\pm\hat{\bm{x}},\pm\hat{\bm{y}}\}$, given by
\begin{subequations}
\begin{align}
    t_{ee}(\bm{n},\bm{n}) ={}&-t_{hh}(\bm{n},\bm{n})= 4t_0-\mu,\\
    t_{eh}(\bm{n},\bm{n}) ={}&t_{he}^\ast(\bm{n},\bm{n})\nonumber\\
    ={}& \tfrac{1}{2}\Delta_0\Bigg[
        \exp\left(-2{i} \int_{\bm{n}}^{\bm{n}+\hat{\bm x}}e\bm A\cdot{d}\bm l + {i}\phi_{{\bm{n}}+\hat{\bm x}} - {i}\phi_{\bm{n}}\right) +
        \exp\left(-2{i} \int_{\bm{n}}^{{\bm{n}}-\hat{\bm x}}e\bm A\cdot{d}\bm l + {i}\phi_{{\bm{n}}-\hat{\bm x}} - {i}\phi_{\bm{n}}\right) \nonumber\\
        &-
        \exp\left(-2{i} \int_{\bm{n}}^{{\bm{n}}+\hat{\bm y}}e\bm A\cdot{d}\bm l + {i}\phi_{{\bm{n}}+\hat{\bm y}} - {i}\phi_{\bm{n}}\right) 
        -  \exp\left(-2{i} \int_{\bm{n}}^{{\bm{n}}-\hat{\bm y}}e\bm A\cdot{d}\bm l + {i}\phi_{{\bm{n}}-\hat{\bm y}} - {i}\phi_{\bm{n}}\right)\Bigg],\\
    t_{ee}(\bm{n}+\hat{\bm \delta},{\bm{n}}) ={}& -t_0 \exp\left({i}\int_{\bm{n}}^{{\bm{n}}+\hat{\bm \delta}}e\bm A\cdot{d}\bm l - {i}\phi_{{\bm{n}}+\hat{\bm \delta}} + {i}\phi_{\bm{n}}\right), \\
    t_{hh}({\bm{n}}+\hat{\bm \delta},{\bm{n}}) ={}& -t_0 \exp\left(-{i}\int_{\bm{n}}^{{\bm{n}}+\hat{\bm \delta}}e\bm A\cdot{d}\bm l\right) ,\\
    t_{eh}({\bm{n}}+\hat{\bm \delta},{\bm{n}}) ={}&t_{he}^\ast({\bm{n}},{\bm{n}}+\hat{\bm \delta})\nonumber\\
    ={}& \tfrac{1}{2}\Delta_0\Bigg[\exp\left({i}\int_{\bm{n}}^{{\bm{n}}+\hat{\bm \delta}}e\bm A\cdot{d}\bm l - {i}\phi_{{\bm{n}}+\hat{\bm \delta}} + {i}\phi_{\bm{n}}\right)+ \exp\left(-{i}\int_{\bm{n}}^{{\bm{n}}+\hat{\bm \delta}}e\bm A\cdot{d}\bm l\right)\Bigg]\times
    \begin{cases}
-1&{\rm if}\;\;\hat{\bm{\delta}}=\pm\hat{\bm{x}},\\
+1&{\rm if}\;\;\hat{\bm{\delta}}=\pm\hat{\bm{y}}.
\end{cases} 
\end{align}
\end{subequations}

\end{widetext}

\section{Quasiparticle density profile near the vortex core}
\label{densityprofile}

\begin{figure}[tb]
\centerline{\includegraphics[width=0.9\linewidth]{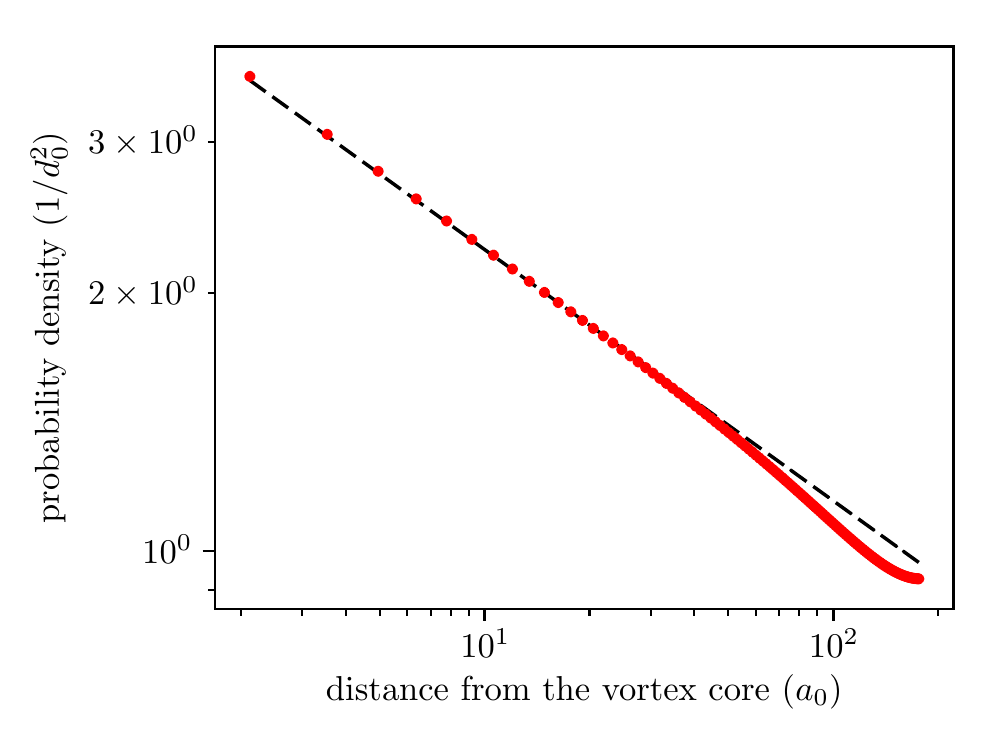}}
\caption{Red data points: Dependence of the probability density $|\psi(x,y)|^2$ on the distance from a vortex core along the line $x=y$, calculated in the zeroth Landau level at momentum $\bm{k}=(\pi/2,\pi/2,\pi/3)$,  for parameters $\Delta_0=1$, $\beta=\sqrt 2$, $\mu=0$, $d_0=502\,a_0$. We took a weaker magnetic field than in Fig.\ \ref{fig_vortexarray} (which had a vortex array with lattice constant $d_0=202\,a_0$), so that the vortices are more widely separated and we can extract the single-vortex asymptotics more easily. The slope of the dashed line is the analytical prediction \eqref{psi2analytical}.
}
\label{fig:asympt}
\end{figure}

In the main text we showed that our numerical simulations reproduce the dispersion relation expected from the analytical theory: The dispersionless zeroth Landau level in the plane perpendicular to the applied magnetic field, see Fig.\ \ref{fig_layout}b, and the linear dispersion along the field, see Fig.\ \ref{fig_kzdispersion}. We also checked that the numerical result $q_{\rm eff}\approx\pm 0.73\,e$ for the effective charge of the quasiparticles at the Weyl point is close to the analytical prediction:
\begin{equation}
|q_{\rm eff}/e|= \sqrt{1-\Delta_0^2/\beta^2}=1/\sqrt 2\approx 0.71.
\end{equation}

As a further test, we compare in Fig.\ \ref{fig:asympt}  the dependence of the quasiparticle density $|\psi|^2$ on the distance $\delta r$ from a vortex core. The analytical prediction from Eq.\ \eqref{smallrasymp}, 
\begin{equation}
|\psi|^2\simeq \delta r^{-1+|q_{\rm eff}|/e}=\delta r^{-1+1/\sqrt 2},\label{psi2analytical}
\end{equation}
is in excellent agreement with the numerics.

\section{Arbitrary angle between internal magnetization and external magnetic field}
\label{arbitraryangle}

\begin{figure*}[tb]
\centerline{\includegraphics[width=0.8\linewidth]{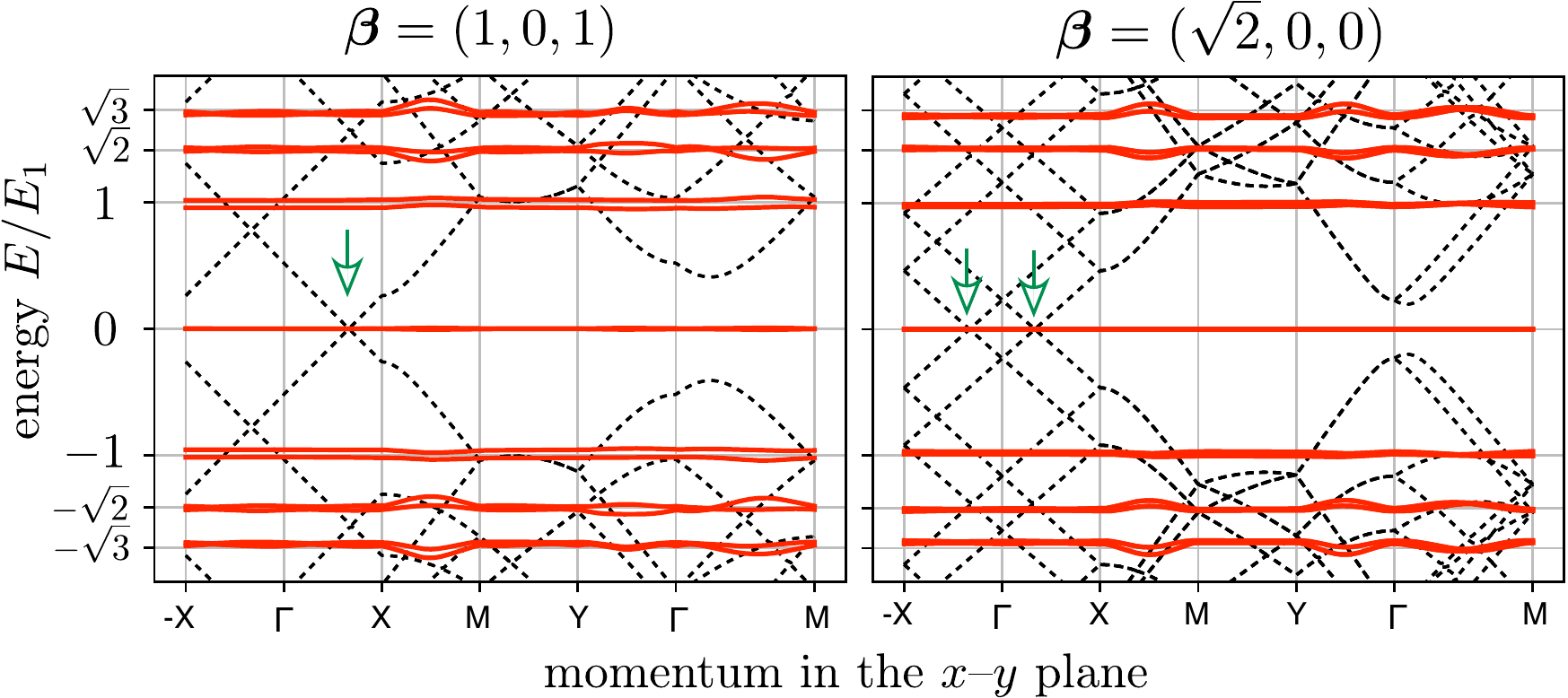}}
\caption{Same as Fig.\ \ref{fig_dispersion}b, but for an internal magnetization $\bm{\beta}$ that is rotated away from the magnetic field $\bm{B}$ in the $z$-direction. The Weyl points are at $\bm{K}=\pm (0.684,0,0.684)$ for $\beta=(1,0,1)$ and at $\bm{K}=\pm(\pi/3,0,0)$ for $\beta=(\sqrt{2},0,0)$, in each case aligned along the magnetization. The $(k_x,k_y)$ momentum is varied along the path through the magnetic Brillouin zone of Fig.\ \ref{fig_layout}b, at fixed $k_z=K_z$, so it passes through one Weyl point for $\beta=(1,0,1)$ and through two Weyl points for $\beta=(\sqrt{2},0,0)$ (green arrows). The flatness of the Landau levels in the vortex lattice is essentially unaffected by the rotation of the magnetization, but the energies themselves are shifted because of the anisotropic Fermi velocity: $E_n=\sqrt n\,E_1$, with $E_1 =(2/d_0)\sqrt{\pi v_{x}v_y}$, and $v_x=1$, $v_y=0.774$ for $\beta=(1,0,1)$; $v_x=1$, $v_y=0.612$ for $\beta=(\sqrt 2,0,0)$.
}
\label{fig_dispersion101}
\end{figure*}

The four-band Hamiltonian \eqref{H0Weyl} of the Weyl semimetal has an internal magnetization $\beta$ pointing in the $z$-direction, parallel to the external magnetic field $\bm{B}=B_0\hat{z}$. If instead the magnetization vector $\bm{\beta}=(\beta_x,\beta_y,\beta_z)$ points in an arbitrary direction, the Hamiltonian becomes
\begin{align}
H_0(\bm{k})={}&t_0{\sum_{\alpha=x,y,z}}\left[\tau_z\sigma_\alpha\sin k_\alpha a_0+\tau_x\sigma_0(1-\cos k_\alpha a_0)\right]\nonumber\\
&+\tau_0\,\bm{\beta}\cdot\bm{\sigma}-\mu\tau_0\sigma_0.\label{H0Weylvector}
\end{align}
Numerical results for the spectrum are shown in Fig.\ \ref{fig_dispersion101} for a magnetization at a $45^\circ$ degree angle and at a $90^\circ$ angle with the magnetic field. The zeroth Landau level remains dispersionless in the $x$--$y$ plane.

We note that now the Weyl cone is anisotropic in the $x$--$y$ plane, but that also does not spoil the protection of the zeroth Landau level.

\section{Tilting of the Weyl cones}
\label{tilted}

\begin{figure*}[tb]
\centerline{\includegraphics[width=0.8\linewidth]{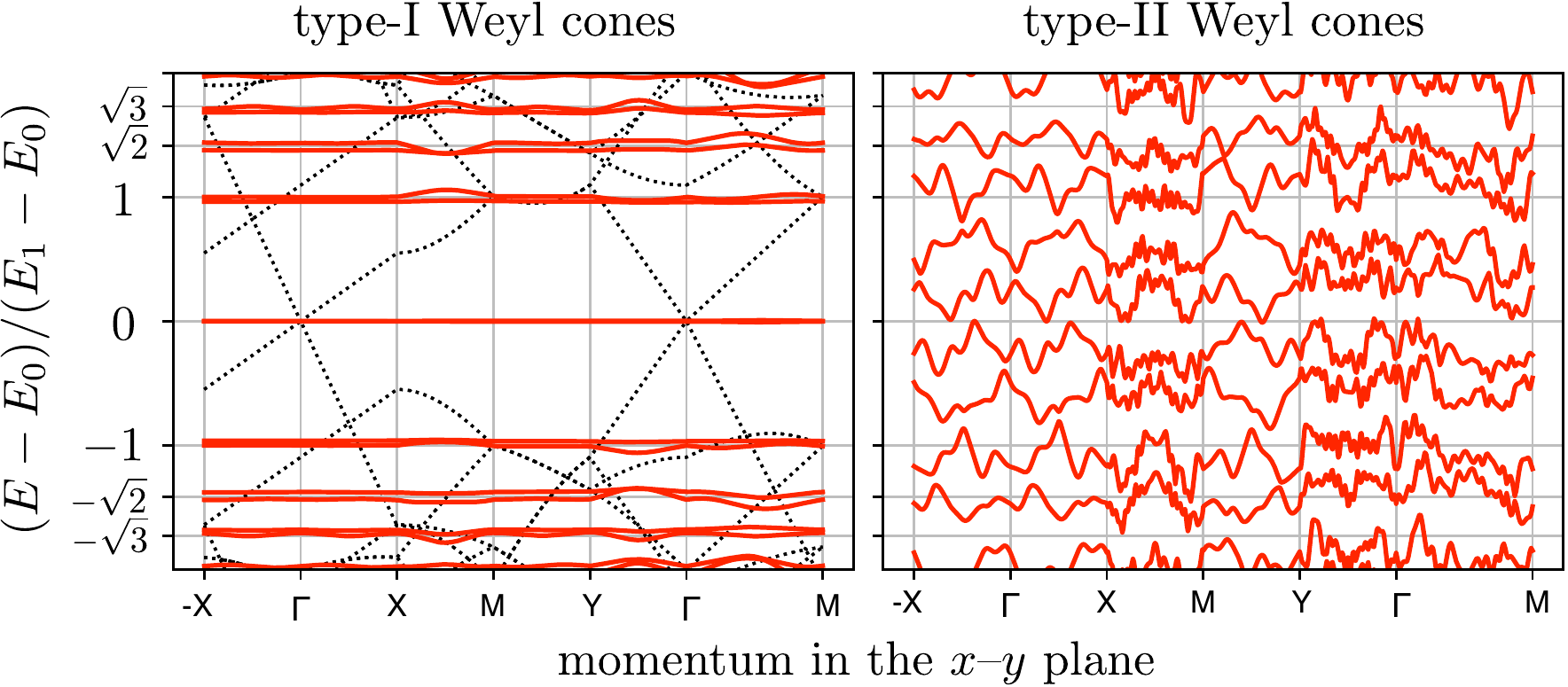}}
\caption{Same as Fig.\ \ref{fig_dispersion}b, but for tilted Weyl cones with $\bm\eta= (0.5,\,0,\,0.05)$ (left panel) and $\bm\eta = (1.1,\,0,\,0)$ (right panel). The energies are shifted by $E_0=-v_{\rm F}\eta_z \sin K$. The energy $E_1$ of the first Landau level was calculated numerically. In the type-I regime $\lvert \bm\eta \rvert < 1$ the Landau levels remain intact. For $\lvert \bm\eta \rvert > 1$ the Weyl superconductor goes through a Lifshitz transition to type-II Weyl cones and the Landau levels disappear.}
\label{fig_tilt}
\end{figure*}

To further explore the robustness of the zeroth Landau level, we consider what happens if we break Lorentz invariance by tilting the Weyl cones. Following Ref.\ \onlinecite{Sol15} one distinguishes type-I from type-II Weyl cones, depending on whether the equi-energy contours are closed elliptic (type-I) or open hyperbolic (type-II). In the absence of superconductivity, it is known that the topological protection of the zeroth Landau level persists all the way up to the Lifshitz transition from a type-I to a type-II Weyl semimetal \cite{Kaw11,Kaw12}. Here we show that the same applies in the superconducting vortex lattice.

\subsection{Hamiltonian of a type-I Weyl supserconductor}
 
We break Lorentz-invariance (particle-hole symmetry) of the Hamiltonian \eqref{H0linear} by adding momentum dependent terms proportional to the unit matrix, 
\begin{equation}
H_0(\bm k) =  v_{\rm F}\tau_z \bm k\cdot\bm\sigma + \beta\tau_0\sigma_z - \mu \tau_0\sigma_0  - v_{\rm F}(\bm\eta \cdot\bm k)\tau_0\sigma_0.
\end{equation}
The Weyl cones are tilted in the direction of the vector $\bm{\eta}$. To simplify the equations we orient the $x$--$y$ axes so that the cones are tilted in the $x$--$z$ plane, hence without loss of generality we may set $\eta_y=0$ (allowing for both $\eta_x$ and $\eta_z$ to be nonzero). The equi-energy contours are closed elliptic (type-I Weyl cone) for $|\bm{\eta}|<1$.

The low-energy Hamiltonian, obtained by the unitary transformation \eqref{Vdef} followed by a projection on the 
$\nu=\tau=\pm 1$ subspace, is
\begin{align}
&H_\pm(\bm k) = v_{\rm F}\textstyle{\sum_{\alpha=x,y}}(k_\alpha + a_\alpha \pm\kappa mv_{\rm{s},\alpha})(\sigma_\alpha-\eta_\alpha\sigma_0) \nonumber\\
&\quad+(\beta-m_{k_z})\sigma_z  \mp \kappa\mu\sigma_0- v_{\rm F} k_z \eta_z\sigma_0.
\label{HLEtilt}
\end{align}
For $\lvert k_z\rvert = K$ at the Weyl point, this reduces to 
\begin{equation}
H_\pm(\bm{k})=H_{\rm chiral}+E_\pm\sigma_0,\;\;E_\pm=\mp \kappa\mu- v_{\rm F} K\eta_z,
\end{equation}
where $H_{\rm chiral}$ differs from Eq.\ \eqref{Hchiraldef} by the appearance of diagonal terms,
\begin{equation}
H_{\rm chiral} = v_{\rm F}\begin{pmatrix}
-\eta_x \Pi_x &\Pi_x-i\Pi_y\\
\Pi_x+i\Pi_y & -\eta_x \Pi_x 
\end{pmatrix}.\label{Hchiraltilt}
\end{equation}

\subsection{Generalized chiral symmetry protects the zeroth Landau level}

The Hamiltonian \eqref{Hchiraltilt} no longer anticommutes with $\sigma_z$, so chiral symmetry is broken. However, following Refs.\  \onlinecite{Kaw11,Kaw12}, for $|\eta_x|<1$ we can generalize the chiral symmetry relation by means of the non-Hermitian operator
\begin{equation}
\gamma = \lambda^{-1}\sigma_z(\sigma_0 - \eta_x\sigma_x),\quad\lambda=\sqrt{1-\eta_x^2},
\end{equation}
such that 
\begin{equation}
\gamma^\dag H_{\rm chiral} +H_{\rm chiral}\gamma=0,\;\;\gamma^2=1.\label{generalizedchirality}
\end{equation}

The right eigenvectors of $\gamma$ are 
\begin{equation}
\begin{split}
&\lvert +\rangle = \frac{1}{\sqrt{2+2\lambda}}\begin{pmatrix}
1+\lambda \\ \eta_x
\end{pmatrix},\\
&\lvert -\rangle = \frac{1}{\sqrt{2+2\lambda}}\begin{pmatrix}
\eta_x  \\ 1+\lambda
\end{pmatrix},
\end{split}\label{plusminbasis}
\end{equation}
with $\gamma|\pm\rangle=\pm |\pm\rangle$. The generalized chirality relation \eqref{generalizedchirality} implies that
\begin{equation}
\langle +|H_{\rm chiral}|+\rangle=0=\langle -|H_{\rm chiral}|-\rangle.\label{diagonalzero}
\end{equation}

Upon substitution of $|\psi\rangle=\psi_+|+\rangle+\psi_-|-\rangle$ the zero-mode equation $H|\psi\rangle=0$ takes the form
\begin{equation}
\begin{pmatrix}
0&\tilde{\cal D}\\
\tilde{\cal D}^\dagger&0
\end{pmatrix}\begin{pmatrix}
\psi_+\\
\psi_-
\end{pmatrix}=0,\;\;\tilde{\cal D}=\frac{1}{v_{\rm F}\lambda}\langle +|H_{\rm chiral}|-\rangle.
\end{equation}
The matrix elements on the diagonal vanish in view of Eq.\ \eqref{diagonalzero}.
The off-diagonal term $\tilde{\cal D}$ equals
\begin{equation}
\tilde{\cal D}=\lambda\Pi_x-i\Pi_y.
\end{equation}
This is almost of the form \eqref{Hchiraldef}, except for the factor-$\lambda$ rescaling of $\Pi_x$. If rescale the coordinates as $x'=x/\lambda$, $y'=y$, and the gauge potential as ${\cal A}'_x=\lambda{\cal A}_x$, ${\cal A}'_y={\cal A}_y$, we have equivalently
\begin{equation}
\tilde{\cal D}=\Pi'_{x}-i\Pi'_{y},\;\;\bm{\Pi}'=-i\nabla'+e{\cal A}'.
\end{equation}
The rescaling does not affect the existence of the zeroth Landau level, nor its degeneracy, since the enclosed flux is unchanged:
\begin{align}
\Phi'{}&=\int dx'\int dy'\,(\partial_{x'}{\cal A}'_y-\partial_{y'}{\cal A}'_x)\nonumber\\
&=\int \frac{dx}{\lambda}\int dy\,(\lambda\partial_{x}{\cal A}_y-\partial_{y}\lambda{\cal A}_x)\nonumber\\
&=\int dx\int dy\,(\partial_{x}{\cal A}_y-\partial_{y}{\cal A}_x)=\Phi.
\end{align}

We conclude that the zeroth Landau level remains topologically protected against scattering by the superconducting vortex lattice even if Lorentz invariance is broken by tilting the Weyl cones --- up to the Lifshitz transition at $|\bm\eta|=1$ from type-I to type-II Weyl cones \cite{note_app}. In Fig.\ \ref{fig_tilt} we show numerical data that confirms this conclusion from the analytics.

\subsection{Chiral dispersion along the magnetic field} 

To complete the calculation we examine the dispersion of the zeroth Landau level in the $k_z$-direction, parallel to the magnetic field. We go back to the Hamiltonian \eqref{HLEtilt}, without setting $k_z=K$. In the basis \eqref{plusminbasis} the eigenvalue equation $(H-E)|\psi\rangle=0$ takes the form
\begin{align}
&\begin{pmatrix}
{\cal E}_2+\lambda{\cal E}_1&v_{\rm F}\lambda\tilde{\cal D}+\eta_x{\cal E}_2\\
v_{\rm F}\lambda\tilde{\cal D}^\dagger+\eta_x{\cal E}_2&{\cal E}_2-\lambda{\cal E}_1
\end{pmatrix}\begin{pmatrix}
\psi_+\\
\psi_-
\end{pmatrix}=0,\nonumber\\
&{\cal E}_1=\beta-m_{k_z},\;\;{\cal E}_2=\mp \kappa\mu- v_{\rm F} k_z\eta_z-E.\label{E1E2eq}
\end{align}

We seek a solution
\begin{equation}
\begin{pmatrix}
\psi_+\\
\psi_-
\end{pmatrix}=\begin{pmatrix}
\exp\left(ix\eta_x{\cal E}_1/v_{\rm F}\lambda\right)\phi_+\\
\exp\left(-ix\eta_x{\cal E}_1/v_{\rm F}\lambda\right)\phi_-
\end{pmatrix}\label{psiphirelation}
\end{equation}
with either $\phi_+\equiv 0$ or $\phi_-\equiv 0$. Substitution into Eq.\ \eqref{E1E2eq} gives
\begin{equation}
\begin{split}
\text{either}\;\;&\phi_+\equiv 0\Rightarrow\tilde{\cal D}\phi_-=0\;\;\text{and}\;\;{\cal E}_2=\lambda{\cal E}_1,\\
\text{or}\;\;&\phi_-\equiv 0\Rightarrow\tilde{\cal D}^\dagger\phi_+=0\;\;\text{and}\;\;{\cal E}_2=-\lambda{\cal E}_1.
\end{split}
\end{equation}
The boundary condition \eqref{psiboundary} on the vortex core selects one of these two solutions, depending on the sign of the effective charge $q_{\rm eff}$. 

We conclude that the zeroth Landau level has the $k_z$-dispersion
\begin{equation}
E_\pm(k_z)= ({\rm sign}\,q_{\rm eff})(\lambda\beta-\lambda m_{k_z}) \mp \kappa\mu- v_{\rm F} k_z\eta_z.
\end{equation}
For $\bm{\eta}=0$, $\lambda=1$ we recover the dispersion \eqref{EpmzerothLL} for untilted Weyl cones. The Landau level remains dispersionless in the $k_x$--$k_y$ plane for any $k_z$.


\begin{thebibliography}{99}
\bibitem{note5} We use the heterostructure model of Ref.\ \onlinecite{Men12} for concreteness, but we have checked that the Landau levels appear as well in the model of Ref.\ \onlinecite{Bed15}, which refers to a Weyl semimetal with intrinsic superconductivity. The difference between the two models, scalar versus pseudoscalar pairing \cite{Bov17}, does not affect the topological protection.
\bibitem{note1} A supercurrent perpendicular to the layers can gap out the Weyl cones even if $\Delta_0<\beta$, but for now we only consider supercurrents flowing in the plane of the layers. In the presence of time-reversal symmetry, for $\beta=0$, the Weyl superconductor is gapped except at the vortex cores, so no Landau level can exist in the bulk.
\bibitem{note2} The low-energy Hamiltonian \eqref{HWeyldef} does not include virtual transitions to higher bands, of second order in $\mu$ and $v_s$. These are included in the numerics, which is based on the full Hamiltonian.
\bibitem{note3} To verify Eq.\ \eqref{uvWdef}, first note that $(-i\partial_x\pm \partial_y)f(x\pm iy)=0$, so we only need to consider derivatives of $W$. For that purpose it is helpful to write ${\cal A}=\nabla\times \hat{z}\omega$ (which is possible in the gauge where $\nabla\cdot{\cal A}=0$), then note that ${\cal B}=-\nabla^2\omega$ and use the identity $\nabla^2\ln|\bm{r}-\bm{r}'|=2\pi\delta(\bm{r}-\bm{r}')$, to derive that $W=-(\pi/\Phi_0)\omega=-e\omega$ and $(-i\partial_x\pm \partial_y)W=\mp e{\cal A}_x-ie{\cal A}_y$. Hence one concludes that $(\Pi_x\pm i\Pi_y)e^{\pm W}=0$.
\bibitem{param2} The parameters for the band structure of Fig.\ \ref{fig_kzdispersion} are the same as those of Fig.\ \ref{fig_dispersion}b, except that we took a larger magnetic field ($d_0=10$) so that the Landau level splitting is more clearly visible on this scale.
\bibitem{param3} The wave function in Fig.\ \ref{fig_vortexarray} is evaluated for the same parameters as Fig.\ \ref{fig_dispersion}b, but at a smaller magnetic field ($d_0=202$), to have a smaller overlap of the vortices. There are two zero-modes of opposite effective charge $q_{\rm eff}=\pm\kappa e$, with identical density profile so we only show one of them.
\end{thebibliography}
\begin{thebibliography}{99}
\bibitem{note_app} Because the generalized chiral symmetry of tilted Weyl cones  requires $\eta_x^2+\eta_y^2<1$, irrespective of the tilt $\eta_z$ in the $z$-direction, the protection of the zeroth Landau level for nonzero $\eta_z$ extends somewhat beyond the Lifshitz transition at $\eta_x^2+\eta_y^2+\eta_z^2=1$.
\end{thebibliography}
\end{document}